\documentclass[11pt]{article}

\usepackage[final]{acl}

\usepackage{times}
\usepackage{latexsym}
\usepackage[T1]{fontenc}
\usepackage[utf8]{inputenc}
\usepackage{microtype}
\usepackage{inconsolata}
\usepackage{graphicx}
\usepackage{booktabs}
\usepackage{amsmath}
\usepackage{amssymb}

\newcommand{\benchmark}{SWE-Recreation}

\title{SWE-Recreation: Benchmarking Coding Agents for Software Recreation}

\author{
  Zhiyao Xu\thanks{Equal contribution.} \\
  Tsinghua Shenzhen International \\ Graduate School, Shenzhen, China \\
  \texttt{xu-zy25@mails.tsinghua.edu.cn} \And
  Zhanjie Ding\textsuperscript{*} \\
  Tsinghua Shenzhen International \\ Graduate School, Shenzhen, China \\
  \texttt{dingzj25@mails.tsinghua.edu.cn} \AND
  Hongyao Yu\textsuperscript{*} \\
  Tsinghua University, Beijing, China \\
  \texttt{3075401625@qq.com} \And
  Jingyu Xiao \\
  The Chinese University of Hong Kong, \\ Hong Kong, China \\
  \texttt{jyxiao@link.cuhk.edu.hk}
}

\begin{document}
\maketitle

\begin{abstract}
Agents increasingly operate software and even write it, yet they are almost always told what to achieve: given a goal, a mock-up, or a demonstration. We ask whether an agent can instead understand an unfamiliar product by using it, and then rebuild it. The agent sees only rendered pixels and acts through pointer and keyboard, without access to the DOM, source, logs, or back-end. Within this interface, it explores, generates a replica, and checks its own work before evaluation. We present \benchmark, a benchmark of 54 real applications (28 websites and 26 Android apps) with 1,446 hidden requirements describing multi-step, stateful workflows and explicit exclusions for behavior outside the reconstruction scope. A fixed judge grades workflows with pixel evidence; a separate post-hoc judge estimates exploration coverage. Across 54 evaluated applications, five frontier systems reach at most 0.530 weighted fidelity and 22.9\% fully reproduced requirements under application-equal averaging. These descriptive results expose both exploration-evidence gaps and implementation shortfalls. Code and data are available at https://github.com/nologistic/recreate-anything.

\end{abstract}
\section{Introduction}

Language-model agents are learning to work with software from both sides. On one side, web and computer-use agents carry out long instructions inside live interfaces~\citep{zhou2024webarena,he2024webvoyager,xie2024osworld,deng2023mind2web,qin2025uitars,xu2026spatio,zhou2026st}; on the other, coding agents turn screenshots~\cite{zhao2026beyond,tang2025slidecoder,tang2026efficientpostergen}, specifications, and issues~\cite{xiao2026visualrepair} into working programs~\citep{si2025design2code,lu2025webgenbench,jimenez2024swebench}. Both lines, however, rest on a quiet assumption: the agent is told what to achieve, whether through a stated goal, a supplied mock-up, or a demonstrated behavior. Human developers routinely face a harder situation---an unfamiliar product with no manual---and resolve it by using the product until they understand it well enough to build it themselves. Whether agents can do the same remains largely untested.

\begin{figure}[h]
    \centering
    \includegraphics[width=\linewidth]{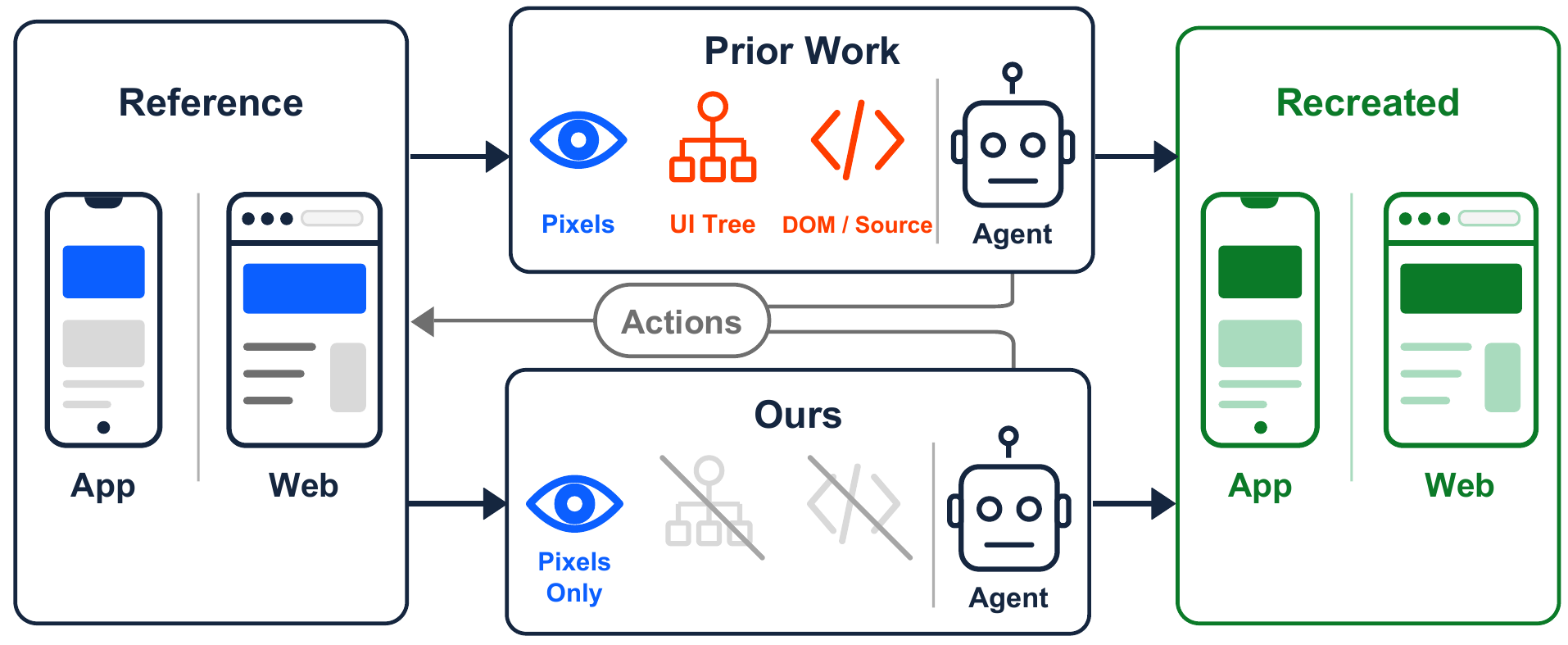}
    \caption{Comparison between \benchmark \ and prior works.}
    \label{fig:intro_compare}
\end{figure}

Recent work has moved toward this setting but stops short in one of two ways. Interactive-reconstruction benchmarks replace the static screenshot with a recorded interaction video or annotated state transitions, then score a replica along a fixed trajectory~\citep{xiao2025interaction2code,chen2026iwrbench,xu2026video2code,dai2026webvr}; the evidence is richer, but someone else has already chosen which behavior to show. Concurrent work lets the agent probe a running reference itself, but through code- and tool-level access to its internals~\citep{recreationworld2026}. The former relieves the agent of deciding what to look at; the latter lets it read the machine rather than the screen. Neither asks whether an agent can come to understand software the way a person does.

We study exactly this question through \emph{active black-box application reconstruction}. An agent is given a running application it has never seen and may engage with it only as a human user can: it sees rendered pixels and acts through pointer and keyboard, with no access to the DOM, source, logs, or back-end. Within this human-like interface, it explores the application, builds a runnable replica, and checks its own work before the result is evaluated. Pixels and pointer input are the one interface every website and application exposes, whereas structured hooks are often missing or unusable in native, canvas-based, remote, or protected front ends; an agent that does without them can, in principle, operate wherever a person can. Withholding these shortcuts is also what makes the task a genuine test of autonomy: with no structured map of the interface and no human-authored hints, the agent must plan its own exploration and decide for itself when it has learned enough to build.


The same freedom that makes the task realistic also makes it hard, in ways that a supplied trajectory would conceal. Because the agent chooses its own observations, failure can enter at three distinct points: it may never \emph{discover} a capability, \emph{misread} what it observed, or \emph{understand} a behavior yet fail to \emph{implement} it. Each risk is aggravated by partial observability, by visual feedback whose cause is ambiguous---the same change on screen may reflect local UI state, shared data, validation, or persistence---and by a horizon that is long twice over, first in the exploration session and then in the stateful workflows to be rebuilt. Success therefore depends on autonomous exploration, sustained long-horizon planning and tool use, and the ability to compile scattered observations into a functionally faithful program.

To measure these abilities, we introduce \textsc{\benchmark}, a benchmark built on real websites and mobile applications. Each target comes with a hidden, human-authored checklist of complete functional workflows. Each requirement specifies a concrete functionality, an ordered sequence of interaction steps, and the expected observable outcomes and state changes upon successful completion---such as filtering and then editing a persistent collection, or creating an object and confirming it across views. An explicit \emph{exclusion} set covers behaviors that no single machine can faithfully reproduce, such as multi-user, payment, real-time, or generative features. A single fixed judge model evaluates each replica through the same human-like interface, but its interaction is \emph{adaptive}: rather than replaying predefined scripts or fixed trajectories, the judge autonomously navigates the replica, exercises each specified workflow, and verifies the resulting behavior against the expected outcomes, citing pixel evidence for its grades. This adaptive evaluation accommodates diverse application structures and behaviorally equivalent implementations without requiring manually tailored evaluation scripts for individual tasks, while distinguishing functional replicas from static mock-ups. Beyond this headline fidelity, our metrics distinguish what an agent \emph{discovered} from what it \emph{implemented} and expose output that merely appears to work, making it possible to locate where, on the path from perception to code, an agent breaks down.

Evaluating five frontier agent systems shows that the task is far from solved: the strongest reaches a combined weighted fidelity of 0.530 and fully reproduces fewer than one requirement in four under application-equal averaging. Kimi and DeepSeek lead both platform strata, but the remaining ordering changes with the application mix. On the web, requirement-level analysis separates low credit from low exploration evidence: automation and recurrence receive the lowest mean credit, browsing and navigation the highest, and mathematical dependencies can fail even when exploration is evidenced. 


Our contributions are threefold:
\begin{itemize}
    \item \textbf{A task.} We formulate \emph{active black-box application reconstruction}, in which an agent must discover an unfamiliar application's functionality and rebuild it through a human-like interface alone, without privileged access or human-authored guidance.
    \item \textbf{A benchmark.} We build \textsc{\benchmark} on 54 real web and mobile applications with 1,446 hidden requirements that stress multi-step state change, cross-view consistency, and persistence, together with an exclusion boundary for behavior that cannot be reproduced on a single machine.
    \item \textbf{An evaluation protocol and study.} We combine functional grading with post-hoc exploration estimates, and provide auditable requirement-level analyses of functional difficulty, exploration evidence, and implementation conditional on evidenced reach.
\end{itemize}
\section{Related Work}
\label{sec:related-work}

\paragraph{Agents that use software.}
Powered by strong multimodal models~\citep{hurst2024gpt4o,bai2025qwen25vl}, agents increasingly operate real software through its interface. Web agents follow instructions across live and reproducible websites~\citep{deng2023mind2web,zhou2024webarena,koh2024visualwebarena,he2024webvoyager,furuta2023multimodal}, computer-use agents extend this to the desktop and to long-horizon, cross-application tasks~\citep{xie2024osworld,yuan2026osworld2}, and parallel efforts target mobile applications and general assistants~\citep{rawles2024androidworld,mialon2024gaia}. A supporting line builds the perception such control needs---screen parsing, web-page grounding, and native GUI action models~\citep{lee2023pix2struct,liu2024visualwebbench,wu2024osatlas,qin2025uitars}---typically organized around reason-and-act loops~\citep{yao2023react}. In all of these, the agent is handed a goal and judged on whether it can \emph{execute} that goal in an interface that already exists; it is never asked to recover, and then reproduce, the interface's underlying specification.

\paragraph{Agents that build software.}
A complementary line asks agents to produce software, but from an explicit specification. UI-to-code work turns a screenshot or design into front-end code and scores it against the original~\citep{beltramelli2017pix2code,laurencon2024websight,si2025design2code,yun2024web2code,gui2025webcode2m,guo2024iwbench,wan2024mrweb,xiao2025designbench,jiang2025screencoder,yang2025ui2coden,xiao2025efficientuicoder,xiao2026component,le2026uibenchkit,dang2025envisioning,zhang2026animweb}; website- and application-generation benchmarks start from natural-language requirements or product briefs~\citep{lu2025webgenbench,he2026vision2web,tran2026vibecodebench,ran2025appforge}; and coding agents resolve issues or implement features in existing repositories~\citep{jimenez2024swebench,yang2024sweagent,xia2025agentless,wang2025openhands,yang2024swebenchmm,yao2025taubench,jain2025livecodebench}. Across these settings the target is supplied---as a screenshot, a prototype, a requirement document, or an issue---so the challenge is faithful implementation of a known specification. In ours the specification is withheld and hidden inside a running application, so an agent must first discover it by interacting and only then implement it.

\paragraph{Reconstruction from observation, and our setting.}
Closest to us is reconstruction from \emph{observed behavior}. Interaction2Code~\citep{xiao2025interaction2code} and InterWebCoder~\citep{xiao2026interwebcoder} condition on annotated state transitions, while IWR-Bench, Video2Code, and WebVR supply a recorded interaction video and score the replica along a fixed trajectory or visual rubric~\citep{chen2026iwrbench,xu2026video2code,dai2026webvr}; in each, the observations have already been chosen for the model. RecreationWorld instead lets an agent probe a \emph{running} reference, but through code- and tool-level access~\citep{recreationworld2026}, in line with a broader push toward scalable agent environments~\citep{aggarwal2026gymanything}. \textsc{\benchmark} differs in that the agent works solely through a human-like interface and receives no human-authored scaffolding, so exploration and planning are themselves under test; our metrics further separate discovery from implementation. For grading, we build on LLM- and agent-as-judge evaluation~\citep{zheng2023judging,zhuge2024agentjudge,gou2025mind2web2}, with a single judge agent with a consistent model and configuration whose every grade cites pixel evidence.

\begin{figure*}[t]
    \centering
    \includegraphics[width=\linewidth]{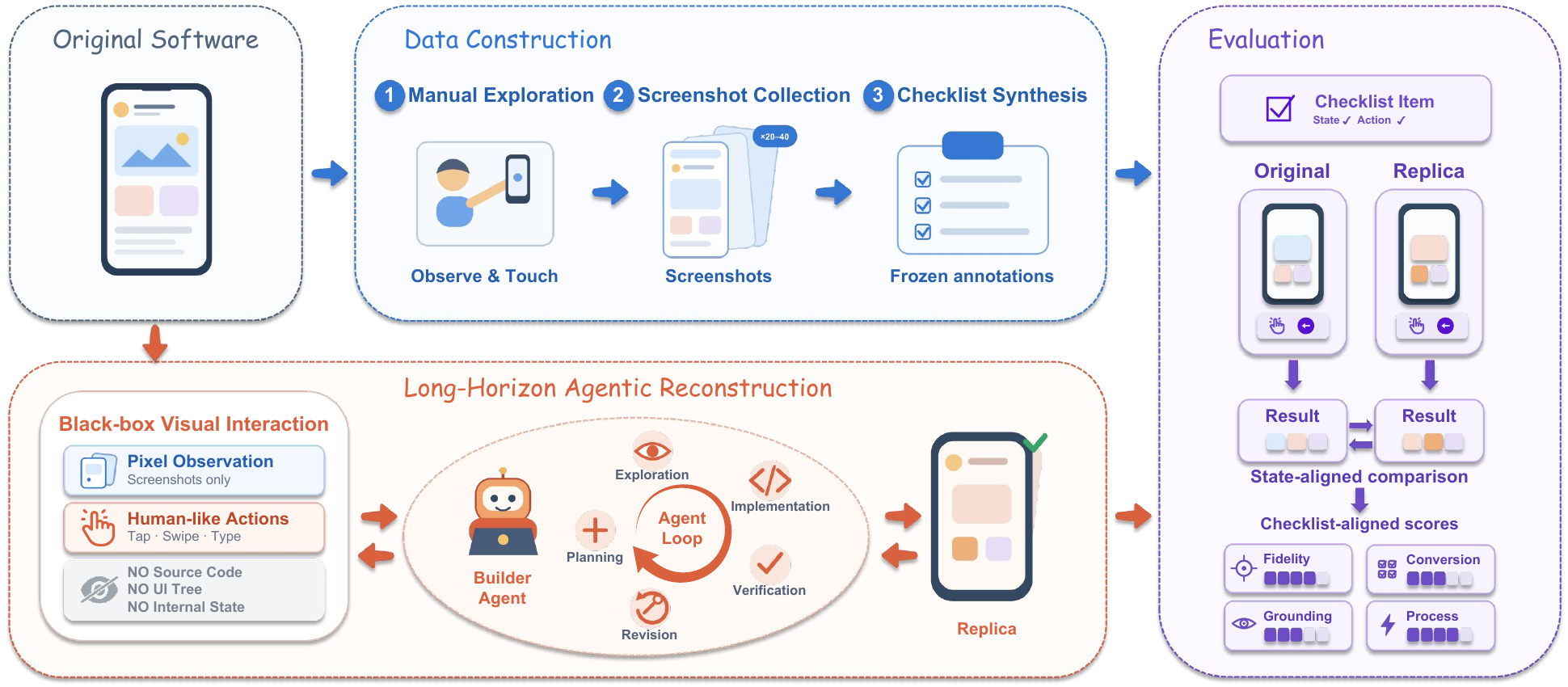}
    \caption{Overview of the \benchmark \ pipeline.}
    \label{fig:pipeline}
\end{figure*}

\section{Task Formulation}
\label{sec:task}

Let $A^\star$ be a source application and $W=\{w_1,\dots,w_N\}$ a hidden set of workflows that a faithful replica must support. An agent's run typically comprises four parts: it \textbf{explores} $A^\star$, \textbf{generates} a runnable replica $\hat{A}$, and \textbf{self-checks} it, after which an \textbf{evaluation} stage grades $\hat{A}$ against $W$ (Section~\ref{sec:eval}). $W$ is never revealed to the agent.

\paragraph{Human-like interface.}
At every step the agent observes only a rendered screenshot and the cursor position, and acts only through coordinate-level pointer and keyboard input (click, drag, scroll, and type on the web; tap, swipe, type, and back on mobile). It receives no DOM, accessibility tree, source code, network traffic, storage, URLs, or selectors, and action receipts report only whether an action was accepted, never what it hit. This boundary is enforced by the runtime rather than by instruction.

\paragraph{Autonomy.}
Exploration is open-ended. No task list, demonstration, exploration plan, or note-taking format is provided, and only a generous, effectively non-binding cap on actions and time is imposed; what to probe, and when enough has been learned, is left entirely to the agent.

\paragraph{Output and scope.}
The replica is built from what the agent observed, together with public fictional assets (Section~\ref{sec:construction}), and is judged solely by observable behavior: whether each $w_i$, including its state changes and persistence, can be carried out on screen. Layout, wording, styling, and framework are out of scope, so behaviorally equivalent replicas pass while static mock-ups do not. Behavior that cannot be faithfully reproduced or verified on a single machine---multi-user collaboration, payments, real-time data, generative features, and external integrations---forms an \emph{exclusion} set $X$ that is neither expected of the agent nor scored.

\section{Benchmark Construction}
\label{sec:construction}

\paragraph{Targets.}
\textsc{\benchmark} covers 54 real applications: 28 live websites and 26 Android apps, grouped into six categories (Table~\ref{tab:stats}). The web split spans social and content platforms (e.g., Zhihu, Reddit, YouTube), productivity and collaboration tools (e.g., Notion, Trello, Google Calendar), commerce and travel services, creative canvases (e.g., Excalidraw, diagrams.net, JS~Paint), and math, map, and media applications (e.g., Desmos, Google Maps, Spotify); the mobile split adds open-source productivity, media, utility, and game apps (e.g., Joplin, VLC, AnkiDroid, Minesweeper). We select applications rich in stateful, long-horizon behavior---filtering, editing persistent objects, and keeping one record consistent across views---rather than static presentation. We discuss the resulting non-determinism and pretraining exposure in Section~\ref{sec:limitations}.

\paragraph{Checklists and exclusions.}
Checklists are built with a human-in-the-loop pipeline. An annotator exercises each application by hand and records 20--40 screenshots covering its main features and transitions; a vision-language model (\texttt{gpt-6-astra}) drafts a checklist from this trace, and a second annotator verifies it against the live application, removing ambiguous or untestable items. Each requirement specifies ordered steps, an expected outcome written in natural language rather than as a mechanical assertion, and flags for persistence and multi-user involvement. Web targets additionally carry an annotator-authored exclusion list recording each excluded surface, its treatment, and a reason; a single file defines both, so exploration and evaluation share the same boundary.

\paragraph{Environments.}
Mobile targets run in a containerized Android system with the application preinstalled, and web targets in a locked-down browser. In both, an MCP server is the agent's only bridge to the device, exposing primitive tools (screenshot, tap or click, swipe or scroll, type) while hiding all device-level mechanisms. For generation, the agent receives a fixed scaffold and an asset package of shared public assets plus fictional per-target content, so no real user data is needed. Network access is permitted but not required, and no web technology stack is prescribed; mobile replicas target a fixed native stack to yield an installable app.

\paragraph{Statistics.}
Table~\ref{tab:stats} and Figure~\ref{fig:dataset_composition} summarize the dataset. The 54 applications carry 1,446 requirements in total, about 30 per website and 23 per mobile app, and roughly one in five requires persistence across a reload or restart. The 438 web exclusions comprise 213 hard, 113 default, and 96 conditional exclusions, plus 16 conditional inclusions that keep a narrowly testable part of an otherwise excluded surface; mobile targets are offline and single-user by construction and need none.

\begin{table}[h]
  \centering
  \small
  \setlength{\tabcolsep}{4pt}
  \begin{tabular}{@{}lcc@{}}
    \toprule
    \textbf{Statistic} & \textbf{Web} & \textbf{Mobile} \\
    \midrule
    Applications                     & 28 & 26 \\
    Requirements                     & 848 & 598 \\
    \;\; per application, mean       & 30.3 & 23.0 \\
    \;\; per application, range      & 12--57 & 13--77 \\
    Persistence requirements         & 143 (16.9\%) & 135 (22.6\%) \\
    Exclusion entries                & 438 & -- \\
    \midrule
    \multicolumn{3}{@{}l}{\textit{Applications by category}} \\
    Productivity \& documents        & 8 & 7 \\
    Social \& content                & 7 & -- \\
    Creative \& media                & 6 & 7 \\
    Commerce \& travel               & 3 & -- \\
    Math, maps \& utilities          & 4 & 7 \\
    Games \& puzzles                 & -- & 5 \\
    \bottomrule
  \end{tabular}
  \caption{\textsc{\benchmark} dataset statistics.}
  \label{tab:stats}
\end{table}

\begin{figure*}[t]
    \centering
    \includegraphics[width=.8\linewidth]{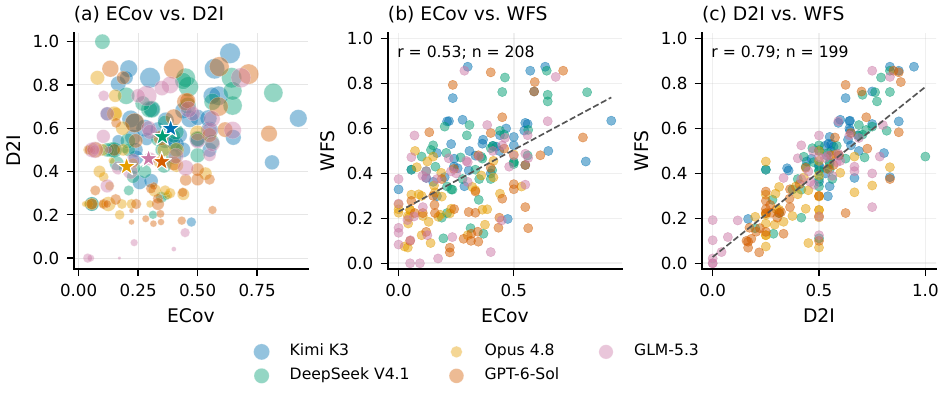}
    \caption{Web and Android session-level associations. (a) Bubble area: WFS; stars: paired means; colors: systems. (b--c) WFS associations with descriptive fits and pair counts. D2I shares grades with WFS; correlations are not causal and platform mixes differ.}
    \label{fig:discovery_implementation}
\end{figure*}

\section{Evaluation Protocol and Metrics}
\label{sec:eval}

\subsection{Protocol}
The judge exercises each hidden workflow $w_i$ on the replica $\hat{A}$ through the human-like interface and compares its behavior with a reference instance. Adaptive functional testing admits behaviorally equivalent implementations while exposing inert interfaces.

A code layer enforces cited observation frames, a closed grade vocabulary, complete requirement coverage, and a fixed report schema; semantic decisions remain with the judge. One fixed model grades every system, avoiding evaluator changes and self-evaluation without eliminating judge bias. Persistence requirements need before, changed, and post-reload/restart evidence for full credit.

\subsection{Notation}
\label{sec:metrics-notation}
For a given target, let $\mathcal{R}$ be its set of scored requirements, with $N=|\mathcal{R}|$ after exclusions are removed, and let $\mathcal{R}_{\mathrm{p}}\subseteq\mathcal{R}$ be those flagged for persistence. For each $r\in\mathcal{R}$, the judge returns a grade $g_r\in\mathcal{G}=\{\textsc{full},\textsc{partial},\textsc{placeholder},\textsc{broken}\}$, a rationale $\rho_r$, and a set $E_r$ of replica screenshots cited as evidence. A separate coverage judge, run post hoc on exploration keyframes without affecting the agent, returns $c_r\in\{0,1\}$ for requirements with an available coverage judgment, and an evidence frame $f_r$ for positive judgments. Here $c_r=0$ means \emph{not evidenced as reached}, not verified absence of exploration; missing judgments are undefined. Positive judgments may evidence an entry point or partial workflow rather than complete execution. Let $\mathcal{O}\subseteq\mathcal{R}$ denote requirements with available judgments. Grades are converted to credit by
\begin{equation}
\omega(g)=
\begin{cases}
1, & g=\textsc{full},\\
0.5, & g=\textsc{partial},\\
0.25, & g=\textsc{placeholder},\\
0, & g=\textsc{broken},
\end{cases}
\label{eq:omega}
\end{equation}


\subsection{Metrics}
\label{sec:metrics-primary}
We group metrics by question: \textbf{Fidelity} assesses functional outcomes; \textbf{Conversion} relates exploration evidence to implementation; \textbf{Grounding} describes visual and interaction proxies; \textbf{Process} records exploration effort, self-checking, and platform reliability.

\subsubsection{Fidelity}
\textbf{Weighted Fidelity Score (WFS)} aggregates requirement credit:
\begin{equation}
  \mathrm{WFS}=\frac{1}{N}\sum_{r\in\mathcal{R}}\omega(g_r).
  \label{eq:wfs}
\end{equation}

\textbf{Grade shares (F/P/Pl/B)}, $s_g=\frac{1}{N}\sum_{r\in\mathcal{R}}\mathbf{1}[g_r=g]$ for $g\in\mathcal{G}$, expose its composition. 

\textbf{Persistence Fidelity (P-WFS)} restricts credit to persistence requirements:
\begin{equation}
  \mathrm{P\text{-}WFS}=\frac{1}{|\mathcal{R}_{\mathrm{p}}|}\sum_{r\in\mathcal{R}_{\mathrm{p}}}\omega(g_r),
\end{equation}
It is undefined when $\mathcal{R}_{\mathrm{p}}$ is empty.

\textbf{Nonfunctional Rate (NFR)} is the fraction graded \textsc{placeholder} or \textsc{broken}, comparable across platforms. The supplementary web \textbf{Illusion Rate (IR)} additionally flags rationales describing fake, decorative, or inert behavior via $\psi(\rho_r)$, with $\mathcal{G}_{\times}=\{\textsc{placeholder},\textsc{broken}\}$:
\begin{equation}
  \mathrm{IR}=\frac{1}{N}\sum_{r\in\mathcal{R}}\mathbf{1}\bigl[g_r\in\mathcal{G}_{\times}\lor\psi(\rho_r)\bigr]
\end{equation}

\subsubsection{Conversion}
\textbf{Exploration Coverage (ECov)} is the fraction evidenced as reached among available judgments:
\begin{equation}
  \mathrm{ECov}=\frac{1}{|\mathcal{O}|}\sum_{r\in\mathcal{O}}c_r,
\end{equation}
It is undefined for empty $\mathcal{O}$; missing judgments are not zeros.

\textbf{Discovery-to-Implementation Rate (D2I)} is mean credit conditional on evidenced reach:
\begin{equation}
  \mathrm{D2I}=\frac{\sum_{r\in\mathcal{O}}c_r\,\omega(g_r)}{\sum_{r\in\mathcal{O}}c_r},
  \label{eq:d2i}
\end{equation}
D2I is undefined without positive reach judgments. These metrics do not attribute failures causally; WFS need not equal their product because unreached or unobserved requirements can earn credit.

\subsubsection{Grounding}
\label{sec:metrics-diag}
For $T$ exploration actions and $M$ observations, let $\delta_m$ be consecutive-frame pixel change and $\varepsilon=0.002$. \textbf{Frames}, $\sum_m\mathbf{1}[\delta_m>\varepsilon]$, counts changed observations, not globally unique states. \textbf{Effective- and invalid-action rates (EAR/IAR)} are
\begin{equation}
  \mathrm{EAR}=\frac{1}{M}\sum_{m=1}^{M}\mathbf{1}[\delta_m>\varepsilon],
\end{equation}
\begin{equation}
  \mathrm{IAR}=\frac{1}{T}\sum_{t=1}^{T}\iota_t,
\end{equation}
where $\iota_t=1$ for rejected or error-raising actions. EAR uses observations as its denominator; visual change need not indicate useful exploration, nor does accepted input guarantee correct targeting.

\textbf{Coverage-discounted Visual Similarity (VSim)} uses $\ell_2$-normalized CLIP ViT-L/14 embeddings $\phi$~\citep{radford2021clip} and $b_r=\mathbf{1}[g_r\neq\textsc{broken}]$:
\begin{equation}
  v_r=c_r\,b_r\max_{e\in E_r}\phi(f_r)^{\top}\phi(e),
\end{equation}
with an empty maximum set to $0$ and $\mathrm{VSim}=\frac{1}{N}\sum_{r\in\mathcal{R}}v_r$. The exported scorer assigns zero without an exploration anchor, unlike ECov/D2I's exclusion of missing judgments. We retain these scores without recomputing embeddings; this coverage- and grade-gated measure is not pure perceptual accuracy.

\subsubsection{Process}
\textbf{Steps} and \textbf{Time} count exploration actions and wall-clock minutes, measuring exploration interaction cost only. \textbf{Self-correction rounds (SCR)} count recorded replica revisions after the initial review. \textbf{Platform-attributed failures (PAF)} count crashes plus environment-degrading actions. Platform defects are repaired and rerun; failures with a correctly functioning platform remain capability outcomes. PAF describes evaluation reliability, not agent ability.
\section{Experiments}
\label{sec:experiments}

\subsection{Setup}
We evaluate five model--framework pairs (Table~\ref{tab:main}) end to end under the same interface, assets, and effectively non-binding interaction cap, without task hints or privileged access. The fixed judge, GPT-6-Astra, is not among the evaluated systems. Scores characterize whole systems, including their planning, tool use, and context management.
The balanced study covers 16 common websites and all 26 Android apps: 42 targets, 1,111 requirements, 210 runs and 5,555 grades, once per system and target. The benchmark itself remains 54 applications with 1,446 requirements. Targets receive equal weight, so complete combined means weight the web and Android strata 16:26, not 1:1. Three additional Opus web results are retained separately. Missing diagnostics are excluded. Explicit Android task-key joins extend D2I to both platforms; semantic taxonomy remains web-only without a verified Android specification-ID mapping.

\subsection{Human validation of the LLM judge.}
Four GUI-domain experts with master's- or doctoral-level training
manually review the LLM judge's assessments by selecting 20 samples. More than 95\% of the
reviewed judgments were deemed correct by the experts, providing
evidence for the reliability of the automated evaluator on the
reviewed cases.

\subsection{Fidelity}
\begin{figure}[h]
    \centering
    \includegraphics[width=.8\linewidth]{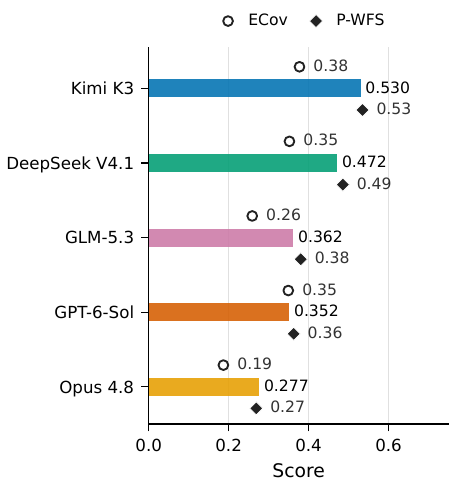}
    \caption{Combined WFS (bars), ECov (circles), and P-WFS (diamonds); denominators follow Table~\ref{tab:main}.}
    \label{fig:system_overview}
\end{figure}

\begin{figure}[h]
    \centering
    \includegraphics[width=\linewidth]{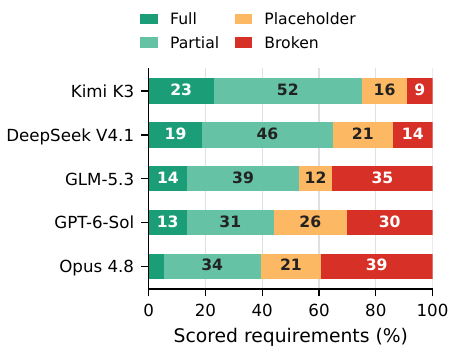}
    \caption{Application-macro grade shares (\%); WFS weights are 1/0.5/0.25/0 for full/partial/placeholder/broken.}
    \label{fig:grade_composition}
\end{figure}

Table~\ref{tab:main} reports combined application-macro results. Kimi leads at 0.530 WFS and 22.9\% full credit, followed by DeepSeek (0.472), GLM (0.362), GPT-6-Sol (0.352), and Opus (0.277). Partial grades remain the largest share for the two leaders (Figures~\ref{fig:system_overview}--\ref{fig:grade_composition}). P-WFS uses a different subset, so its proximity to WFS does not establish that working behavior generally survives reload.

\begin{table*}[t]
\centering\small
\setlength{\tabcolsep}{3pt}
\begin{tabular}{@{}lcccccccccc@{}}
\toprule
 & \multicolumn{7}{c}{\textbf{Fidelity}} & \multicolumn{2}{c}{\textbf{Conversion}} & \multicolumn{1}{c}{\textbf{Grounding}} \\
\cmidrule(lr){2-8} \cmidrule(lr){9-10} \cmidrule(lr){11-11}
\textbf{System} & \textbf{F} & \textbf{P} & \textbf{Pl} & \textbf{B} & \textbf{WFS$\uparrow$} & \textbf{P-WFS$\uparrow$} & \textbf{NFR$\downarrow$} & \textbf{ECov$\uparrow$} & \textbf{D2I$\uparrow$} & \textbf{VSim$\uparrow$} \\
\midrule
Kimi Code + Kimi K3 & \textbf{22.9} & 52.2 & 15.9 & \textbf{9.0} & \textbf{0.530} & \textbf{0.535} & \textbf{0.249} & \textbf{0.377} & \textbf{0.598} & \textbf{0.295} \\
DeepSeek Harness + V4.1-Flash & 18.9 & 45.9 & 21.3 & 13.9 & 0.472 & 0.486 & 0.352 & 0.352 & 0.560 & 0.271 \\
ZCode + GLM-5.3-Flash & 13.5 & 39.4 & 11.7 & 35.3 & 0.362 & 0.380 & 0.471 & 0.259 & 0.459 & 0.191 \\
Codex + GPT-6-Sol & 13.4 & 30.7 & 25.8 & 30.2 & 0.352 & 0.363 & 0.559 & 0.349 & 0.445 & 0.253 \\
Claude Code + Opus 4.8 & 5.2 & 34.3 & 21.1 & 39.3 & 0.277 & 0.269 & 0.605 & 0.187 & 0.423 & 0.118 \\
\bottomrule
\end{tabular}
\caption{Combined application-macro results. F/P/Pl/B: grade shares (\%). WFS, grade shares and NFR use 42 targets; P-WFS uses 38. In row order, ECov/D2I/VSim use 41/40/41, 42/42/42, 42/37/42, 42/42/42, 41/38/41 targets. NFR is grade-only nonfunctionality; other definitions in Section~\ref{sec:metrics-primary}.}
\label{tab:main}
\end{table*}

\noindent Partial credit dominates the strongest replicas, indicating substantial but incomplete functionality rather than consistently faithful workflows.

\subsection{Conversion}
Across platforms, WFS correlates with ECov at $r{=}0.53$ ($n{=}208$) and D2I at $r{=}0.79$ ($n{=}199$), partly because D2I shares its grades (Figure~\ref{fig:discovery_implementation}). These pooled associations also reflect different platform mixes and do not identify causal effects.

\noindent The joint distribution therefore distinguishes implementation shortfalls on observed behavior from gaps in exploration evidence, which aggregate fidelity alone cannot separate.

\subsection{Grounding}
Kimi has the highest combined VSim (0.295) and EAR (0.535), whereas Opus has VSim 0.118 and EAR 0.376 (Tables~\ref{tab:main}--\ref{tab:aux}). These overlapping proxies do not isolate perceptual errors. GPT-6-Sol has higher EAR than DeepSeek (0.507 versus 0.484) but lower WFS (0.352 versus 0.472): visual change alone does not establish useful exploration. All 130 Android visual scores are available after the five pending cases were resolved.

\noindent Read together, the diagnostics reveal different interaction profiles even when individual rates appear favorable; no single proxy adequately summarizes the quality of exploration.

\subsection{Process}
DeepSeek averages 167.9 exploration steps and Kimi 133.9, versus 59.4--102.5 for the other systems (Table~\ref{tab:aux}). Kimi averages 112.9 minutes, compared with DeepSeek's 45.1, so longer action traces need not imply longer elapsed time. Recorded IAR is zero where available; combined PAF is at most 0.143 per target.

\noindent Longer action traces characterize the two leaders, but effort, iteration and fidelity require controlled study before interpreting their association causally.

\begin{table*}[t]
\centering\small
\setlength{\tabcolsep}{3pt}
\begin{tabular}{@{}lccccccc@{}}
\toprule
 & \multicolumn{3}{c}{\textbf{Grounding}} & \multicolumn{4}{c}{\textbf{Process}} \\
\cmidrule(lr){2-4} \cmidrule(lr){5-8}
\textbf{System} & \textbf{Frames} & \textbf{EAR$\uparrow$} & \textbf{IAR$\downarrow$} & \textbf{Steps} & \textbf{Time (min)} & \textbf{SCR} & \textbf{PAF$\downarrow$} \\
\midrule
Kimi Code + Kimi K3 & 119.1 & 0.535 & 0.000 & 133.9 & 112.9 & 2.12 & 0.10 \\
DeepSeek Harness + V4.1-Flash & 138.5 & 0.484 & 0.000 & 167.9 & 45.1 & 2.41 & 0.14 \\
ZCode + GLM-5.3-Flash & 73.3 & 0.399 & 0.000 & 102.5 & 89.5 & 1.26 & 0.05 \\
Codex + GPT-6-Sol & 67.5 & 0.507 & 0.000 & 76.3 & 29.4 & 1.88 & 0.05 \\
Claude Code + Opus 4.8 & 42.0 & 0.376 & 0.000 & 59.4 & 25.2 & 0.17 & 0.02 \\
\bottomrule
\end{tabular}
\caption{Combined available-target process means. Frames/EAR/PAF use 42 targets; IAR uses 41 for GPT-6-Sol and 42 otherwise; Steps/Time use 39 for DeepSeek and 42 otherwise. SCR uses 40, 39, 42, 41, 42 targets in row order and counts revisions after the initial review; definitions in Section~\ref{sec:metrics-primary}.}
\label{tab:aux}
\end{table*}

\subsection{Target Analysis}
\begin{figure}[h]
    \centering
    \includegraphics[width=.8\linewidth]{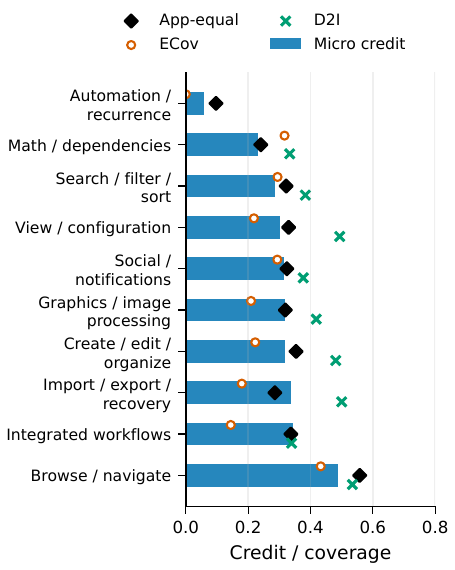}
    \caption{Web functional types: bars show micro credit, diamonds application-equal credit, circles ECov and crosses D2I (absent when undefined). Types are AI-assisted, without independent human validation.}
    \label{fig:req-types}
\end{figure}

Semantic coding of the 513 web requirements finds lowest micro credit for automation/recurrence (0.058) and highest for browsing/navigation (0.487). Automation has no reached pairs, hence undefined D2I; mathematical dependencies score 0.232 despite 31.7\% coverage and D2I 0.333 (Figure~\ref{fig:req-types}). Application-equal averaging preserves these endpoints, not necessarily intermediate rankings (Appendix~\ref{sec:req-analysis}).

Checklist composition limits difficulty claims: 26/36 Trello requirements concern automation, while Desmos Geometry's mean 0.203 and range 0.066 indicate uniformly low scores. Its intersection chain earns zero credit despite 2/5 positive reach judgments; Squoosh's return-and-load workflow earns full credit despite only 2/4. These illustrative contrasts do not identify causes; further requirement-level detail remains in Appendix~\ref{sec:appendix-target-results}.

\noindent Difficulty thus depends on the tested workflows as well as the application, motivating requirement-level reporting alongside application rankings.

\section{Conclusion}
\label{sec:conclusion}

\textsc{\benchmark} tests whether agents can explore and rebuild unfamiliar software through a human-like interface. Five systems reach at most 0.530 combined WFS; Kimi and DeepSeek lead both platform strata, while other rankings depend on the application mix. Grades and joint reach evidence across both platforms separate implementation shortfalls from exploration-evidence gaps. Repeated runs, independent annotation, and controlled interventions are needed to explain these descriptive patterns.

\bibliography{custom}

\appendix
\clearpage

\section{Limitations}
\label{sec:limitations}

The results have several limits. \textbf{Coverage is a post-hoc judge estimate over sampled keyframes}, not ground truth: negative judgments do not verify omissions, and positive judgments may cover only part of a workflow. Missing judgments are excluded rather than counted as non-reach; available-case comparisons can therefore have different system composition. \textbf{VSim is coverage-discounted} and is retained from the existing export, not recomputed from embeddings in this analysis. \textbf{The functional taxonomy is AI-assisted semantic coding}, without independent human annotation or inter-rater reliability. Fifteen flagged source-text inconsistencies motivate a sensitivity analysis; historical grades may themselves reflect these conflicts, and excluding flagged items does not validate all remaining specifications. The checklist-drafting and grading model family may also share biases.

\textbf{Each system has one run per evaluated application}. Requirements within an application are dependent, and functional categories have different application mixes (web automation/recurrence occurs in only two applications). Neither correlations nor category contrasts establish causality. Live targets lack deterministic resets and may have appeared in pretraining. Android exports provide task-level grades and coverage but no rationales, screenshots or verified specification-ID mapping; their persistence aggregates remain supplied values, although ordinal count/score checks agree. Android Frames/SCR use exported counters aligned by the documented metric code, without raw traces for replay. Platform-specific application mixes and runtimes prevent causal platform comparisons. Additional web coverage, repeated runs, independent annotation, and controlled ablations are needed to assess generality and mechanisms.

\section{Web Requirement-level Analysis}
\label{sec:req-analysis}

This analysis asks which behaviors receive credit, how that credit relates to recorded exploration, and how sensitive the resulting comparisons are to checklist composition. The unit of analysis is a requirement--system pair; application and system summaries answer different questions and are kept distinct below.

\subsection{Scope, Joins, and Missing Evidence}
\label{sec:appendix-req-scope}

The analysis uses only the 16 common web targets and the five systems in Table~\ref{tab:main}: 513 distinct requirements, 80 application--system runs, and 2,565 graded requirement--system observations. We join evaluation requirements, review specifications and coverage records by target and requirement ID. One specification uses semantic IDs while its coverage records use numbered IDs; an explicit alias table, checked against the coverage notes for all five systems, resolves that schema difference. Strict checks reject duplicate or unknown IDs and non-Boolean coverage values. Source paths, SHA-256 hashes, full specification text, grade rationales, cited observations and coverage records are exported for audit.

Coverage availability determines the denominator, rather than the implementation grade. There are 2,503 available coverage pairs from 78 runs: 583 reached and 1,920 not evidenced as reached. The other 62 pairs retain their grades but do not enter coverage denominators. Six available runs have zero reached requirements; their ECov is zero and D2I is undefined, leaving 72 defined session-level D2I values. Across requirements, 239 have no positive reach judgment among their available pairs. This is an observed evidence pattern, not proof of omission.

\subsection{Functional Taxonomy and Sensitivity Design}
\label{sec:appendix-taxonomy}

The taxonomy groups requirements by the operation being tested, rather than by the application that contains them.
We assign each requirement one primary functional type by close reading of its title, steps and expected outcome, using an explicit ID-to-type mapping rather than a keyword classifier. This is AI-assisted analysis, not independently validated human annotation. Search/filter/sort covers retrieval and aggregation; create/edit/organize covers domain-record lifecycle; social/notifications covers relationships and communication; view/configuration changes presentation or settings; graphics/image processing changes graphic content; math/dependencies computes mathematical relationships; automation/recurrence executes event/time rules; import/export/recovery transfers or restores state; browse/navigate reads content or moves through it. Integrated workflows explicitly combine multiple operations. Persistence and cross-view consistency remain properties of the source requirement rather than determining its primary type.

Source-text conflicts are treated as uncertainty in the analysis rather than silently repaired specifications. Each exported assignment includes its source text and a semantic justification. Fifteen Squoosh requirements have flagged title/step/expected inconsistencies, including a geospatial-import title for an image-loading workflow and an inventory title for image panning. We code their intended operation using the expected outcome and coherent steps, retain historical grades, and exclude them from illustrative examples. Independent recoding and re-evaluation would be needed to assess annotation validity.

Application-equal sensitivity addresses variation in checklist size.
For a category $k$, let $P_k$ contain its graded requirement--system pairs, $O_k$ those with coverage judgments, and $H_k\subseteq O_k$ those judged reached. We report mean credit $\sum_{p\in P_k}\omega(g_p)/|P_k|$, ECov $|H_k|/|O_k|$, non-reach evidence rate $1-|H_k|/|O_k|$, and D2I $\sum_{p\in H_k}\omega(g_p)/|H_k|$. Empty denominators are undefined. Each requirement has five grades, so micro mean credit equals an equal-weight average of the five-system mean for each requirement. Application-macro sensitivity first computes the same metric within each represented application, pooling available systems, then averages defined application values equally. This differs from the system-specific macro means in Table~\ref{tab:main}; it also prevents large checklists from receiving more weight solely because they contain more requirements. We additionally export results stratified by system.

\subsection{Functional Difficulty and Checklist Composition}
\label{sec:appendix-functional-difficulty}

Figure~\ref{fig:req-types} distinguishes absent reach evidence from low implementation credit despite some evidenced reach. Automation/recurrence has no positive coverage judgments, so its D2I is undefined; math/dependencies has positive coverage but low reached-subset credit. Neither pattern establishes that the full behavior was explored or understood.

The micro bars and application-equal diamonds retain browsing/navigation and automation/recurrence as the highest- and lowest-credit categories, while intermediate positions change. Unequal application mixtures therefore limit any interpretation as a general ordering of intrinsic functional difficulty.

\subsection{Joint Exploration and Implementation States}
\label{sec:appendix-joint-states}

\begin{figure}[h]
    \centering
    \includegraphics[width=\linewidth]{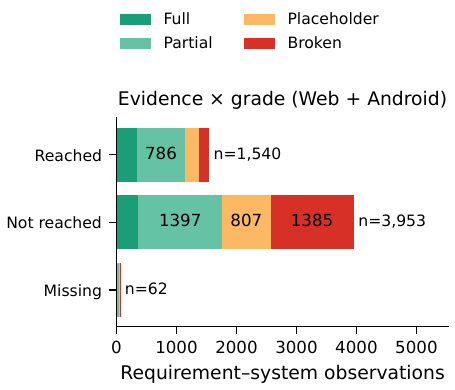}
    \caption{Joint evidence/grade counts over 5,555 requirement--system pairs (2,565 Web and 2,990 Android). Missing coverage is separate from non-reach. ``Not reached'' abbreviates not evidenced as reached.}
    \label{fig:req-states}
\end{figure}

Figure~\ref{fig:req-states} separates reached, non-reached and missing-coverage observations. Both reached and non-reached groups contain full and broken grades, so neither coverage state determines implementation quality. Systems choose what to inspect, and the coverage judge sees sampled keyframes; these distributions do not estimate a causal benefit of exploration.

\section{Dataset Details}
\label{sec:appendix-checklist}
\subsection{Composition and Evaluation Scope}

\begin{figure}[h]
    \centering
    \includegraphics[width=.8\linewidth]{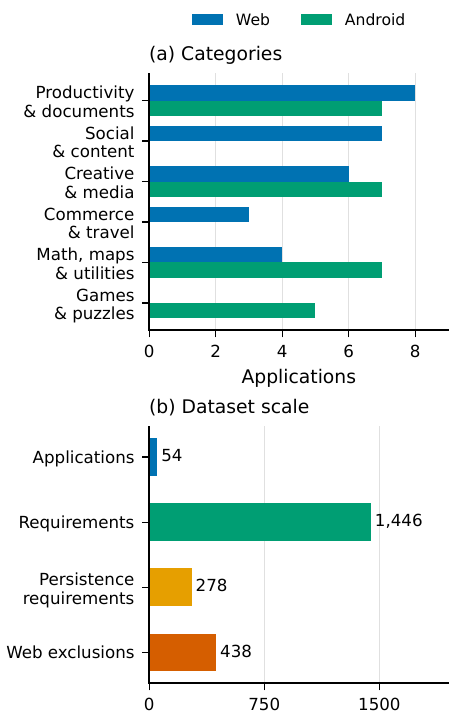}
    \caption{Web/Android application counts by category (a) and benchmark totals (b).}
    \label{fig:dataset_composition}
\end{figure}

The full collection covers complementary application families rather than matched web and mobile versions of the same tasks. Figure~\ref{fig:dataset_composition} shows social/content and commerce/travel targets only on the web, and games/puzzles only on mobile; productivity, creative tools, and utilities occur on both platforms. This composition broadens the behaviors represented by the benchmark, but confounds platform with application family.

The requirement and exclusion counts describe the scope of evaluation, not interchangeable units of difficulty. The collection has 848 web and 598 mobile requirements, including 143 and 135 persistence requirements, respectively (Table~\ref{tab:stats}). The resulting 278 persistence requirements make durable state an explicit part of the benchmark, while remaining a subset of the 1,446 total requirements. The 438 web exclusion entries define the boundary of assessable behavior and are not additional scored tasks. Neither requirement counts nor exclusion counts measure the duration or complexity of a workflow.

\subsection{Application Difficulty and Cross-System Heterogeneity}
\label{sec:appendix-target-results}

\begin{figure}[h]
    \centering
    \includegraphics[width=.8\columnwidth]{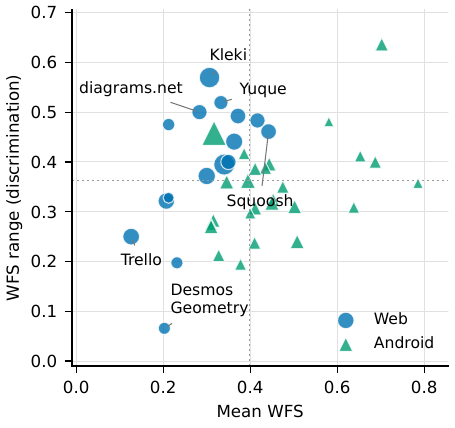}
    \caption{Mean WFS versus cross-system range for web (circles) and Android (triangles); area scales with requirement count. Low mean and narrow range indicate uniformly low scores, not a ceiling task.}
    \label{fig:hardness_discrimination}
\end{figure}

Observed difficulty and cross-system separation are distinct properties of a target. In Figure~\ref{fig:hardness_discrimination}, Desmos Geometry combines a mean of 0.203 with a range of only 0.066, indicating consistently low scores. Among web targets, Kleki has the largest range (0.569), while Squoosh has the highest mean (0.442) but still a range of 0.461. The expanded plot includes Android targets without interpreting platform differences causally. Here ``discrimination'' denotes only observed maximum-minus-minimum WFS, not a calibrated psychometric property or statistical significance.

Checklist composition limits explanations based solely on an application's visible interaction style. Trello's mean of 0.126 and range of 0.250 summarize a checklist dominated by automation requirements, whereas Desmos Geometry's uniformly low scores involve dependency-preserving updates. The plots locate contrasting outcome patterns but do not identify drag-and-drop, mathematical reasoning, or another mechanism as their cause. With one run per system and target, the ranges also provide no estimate of repeat-run variability.

\subsection{Metric Profiles and Their Interpretation}
\label{sec:appendix-profiles}

\begin{figure}[h]
    \centering
    \includegraphics[width=.8\linewidth]{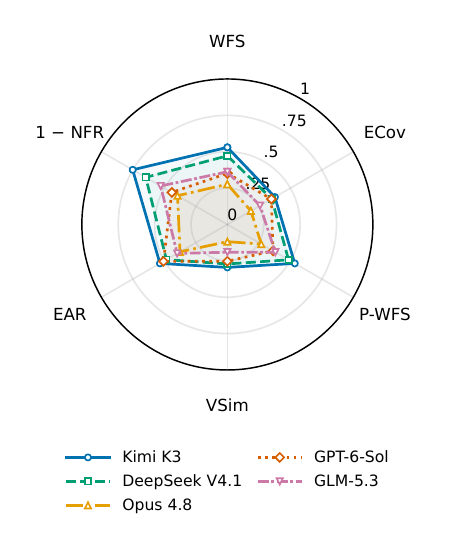}
    \caption{Five system profiles overlaid on shared fixed $[0,1]$ axes using raw metric values; colors, line styles and markers identify systems. Polygon area is not an aggregate score.}
    \label{fig:capability_radar}
\end{figure}

Figure~\ref{fig:capability_radar} overlays the five systems on shared fixed $[0,1]$ axes. The six combined axes use WFS, P-WFS, ECov, VSim, EAR and $1-\mathrm{NFR}$. D2I is shown separately in Figure~\ref{fig:discovery_implementation}; grade-only NFR avoids pooling incompatible rationale-based IR definitions. Fidelity axes reuse grades; ECov and VSim use available targets, and VSim is coverage-discounted. Polygon area is not a validated aggregate capability score and additionally depends on axis ordering; averaging these axes would introduce unsupported weighting and double-count evidence.

\end{document}